\documentclass[final]{svjour2}
\usepackage{graphicx}
\usepackage{rotating}
\usepackage{amssymb}
\usepackage{amsmath}
\usepackage{mathptmx}
\usepackage[numbers]{natbib}
\makeatletter
\journalname{Journal of Low Temperature Physics}

\bibpunct{}{}{,}{s}{}{,}

\begin{document}

\newcommand{\hdblarrow}{H\makebox[0.9ex][l]{$\downdownarrows$}-}
\title{Characterization of the KID-based light detectors of CALDER}

\author{N.~Casali$^{1,2,*}$ \and F.~Bellini$^{1,2}$ \and L.~Cardani$^{1,2,3}$ \and M.G.~Castellano$^{4}$ \and I.~Colantoni$^{1}$ \and A.~Coppolecchia$^{1,2}$ \and C.~Cosmelli$^{1,2}$ \and A.~Cruciani$^{1,2}$ \and A.~D'Addabbo$^{5}$ \and S.~Di~Domizio$^{6,7}$ \and M.~Martinez$^{1,2}$ \and C.~Tomei$^{2}$ \and M.~Vignati$^{2}$}

\institute{$^{1}$ Department of Physics, Sapienza University of Rome, 00185 Piazzale Aldo Moro 2, Rome, Italy \\
$^{2}$ Istisuto Nazionale di Fisica Nucleare - Sezione di Roma, Rome, Italy\\
$^{3}$ Physics Department, Princeton University, Washington Road, 08544, Princeton NJ, USA\\
$^{4}$ Isituto di Fotonica e Nanotecnologie (IFN), CNR, Rome, Italy\\
$^{5}$ Istisuto Nazionale di Fisica Nucleare - Laboratori Nazionali del Gran Sasso, Assergi (AQ), Italy\\
$^{6}$ Department of Physics, University of Genoa, Genoa, Italy\\
$^{7}$ Istituto Nazionale di Fisica Nucleare - Sezione di Genova, Genoa, Italy\\
$^{*}$ Corresponding author, \email{nicola.casali@roma1.infn.it}\\}


\maketitle

\begin{abstract}

The aim of the CALDER (Cryogenic wide-Area Light Detectors with Excellent Resolution) project is the development of light detectors with active area of 5x5~cm$^{2}$ and noise energy resolution smaller than 20 eV RMS, implementing phonon-mediated Kinetic Inductance Detectors (KIDs).\\
The detectors are developed to improve the background suppression in large-mass bolometric experiments such as CUORE, via the double read-out of the light and the heat released by particles interacting in the bolometers. In this work we present the characterization of the first light detectors developed by CALDER.
We describe the analysis tools to evaluate the resonator parameters (resonant frequency and quality factors) taking into account simultaneously all the resonance distortions introduced by the read out chain (as the feed-line impedance and its mismatch) and by the power stored in the resonator itself. We detail the method for the selection of the optimal point for the detector operation (maximizing the signal-to-noise ratio). Finally, we present the response of the detector to optical pulses in the energy range $0-30$~keV.


\end{abstract}

\section{Introduction}\label{secintro}
The CALDER project~\cite{Battistelli:2015vha} aims to develop wide area light detectors (25~cm$^{2}$) with noise energy resolution $< 20$~eV by implementing phonon-mediated Kinetic Inductance Detectors (KIDs)~\cite{Swenson:2010, Moore:2012}. The objective of the first phase of the project is the production of a 2x2~cm$^2$ Si substrate sampled by four 40~nm thick Al film lumped-element resonators as described in Ref.~\cite{Ivan,Ivan2}, with an energy resolution of about 100~eV.

The detectors are operated in a $^3$He/$^4$He dilution refrigerator with base temperature of 10~mK and are permanently exposed to a $^{57}\rm{Co}$ source. With a 400~nm LED light source and an optical fiber that drives the light signals to the detector substrate it is also possible to illuminate the detectors with optical pulses. The output signals that excited the resonators are fed into a CITLF4 SiGe cryogenic low noise amplifier~\cite{ampli} operated at 4~K. A detailed description of the chip design, the cryogenic setup, the room temperature electronics and the acquisition software can be found in Refs~\cite{Battistelli:2015vha, Bour:2011, Bour:2013}. The acquired signals are digitized and saved on disk for the off-line analysis.

\section{Resonance parameters evaluation}\label{secResPar}
The first step of the detectors characterization is the evaluation of the resonators parameters: the resonant frequencies ($f_{0}$) and the internal ($Q_i$) and coupling ($Q_c$) quality factors. The knowledge of these quantities is fundamental not only for the resonator characterization but also for the detector response study and its absolute energy calibration as described in Ref.~\cite{Cardani:2015}.
The transmitted signal ($S_{21}$) for an ideal KID can be written as:
\begin{equation}\label{eqtrans}
S_{21}\left(f\right) = 1-\frac{\frac{Q}{Q_c}}{1+2jQ\cdot \frac{f-f_0}{f_0}}
\end{equation}
where $Q = \left(\frac{1}{Q_i}+\frac{1}{Q_c}\right)^{-1}$ is the total quality factor of the resonator. At frequencies $f\ll f_{0}$ or $f\gg f_{0}$, $S_{21}$ is close to 1 and the feed-line is unaffected by the resonator. When $f$ sweeps through the resonance, $S_{21}$ traces out a circle in the In-phase Quadrature (IQ) plane with diameter equal to $Q/Q_c$ which is referred to as the resonance circle: at the resonant frequency $f = f_0$ the circle crosses the real axis at the closest approach to the origin, $S_{21} = 1-Q/Q_{c}$. In this ideal case the resonance parameters evaluation is quite simple.

Nevertheless, in real measurements there are deviations from the ideal response caused by the read-out chain, by the impedance mismatch between the feed-line and the chip and the power dissipated on the resonator that lead to a translation, rotation, magnification and distortion of the resonance circle, which complicates the evaluation of $f_{0}$, $Q_{i}$, $Q_{c}$ and $Q$:
\begin{itemize}
\item the read-out chain introduces an electronic time delay $\tau$ and a distortion in the IQ plane introduced by the IQ-Mixer, which is in the most general case an ellipse arc with axis $A$ and $B$ and center $Z_c$, rotated by an angle $\phi$;
\item the impedance mismatches in proximity of the KIDs produce a rotation and magnification of the resonance circle that can be represented using only one parameter $\theta$~\cite{Khalil:2012};
\item the power absorbed by the resonator introduces an asymmetry in the resonance shape and a left shift in the resonant frequency $f_{0}$. It is possible to account for this effect performing the substitution $Q\frac{f-f_0}{f_0} = y$ in Eq.~\ref{eqtrans}. The new variable $y$ can be written as $y = y_0+\frac{a}{1+4y^2}$, where $y_0 = Q\left(\frac{f-f_0}{f_0}\right)$, and the parameter $a$ quantify the resonance asymmetry (see Ref.~\cite{Swenson:2013jsa} for more details).
\end{itemize}
\begin{figure}
\begin{center}
\includegraphics[%
  width=1.\linewidth,
  keepaspectratio]{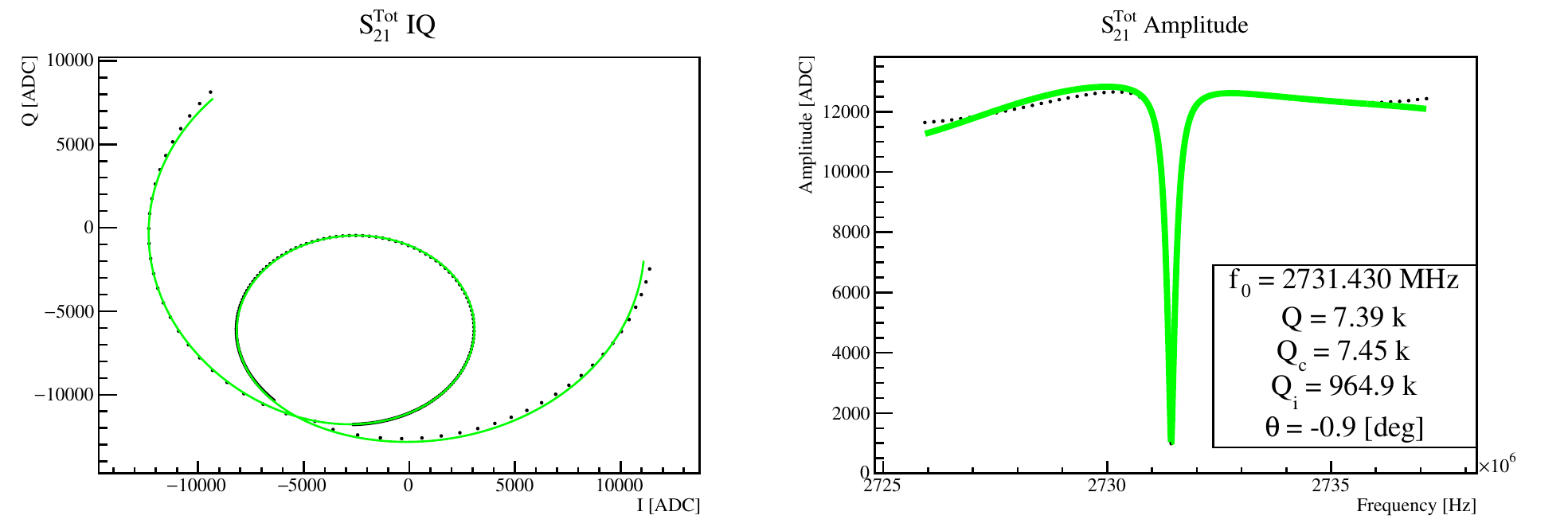}
\includegraphics[%
  width=1.\linewidth,
  keepaspectratio]{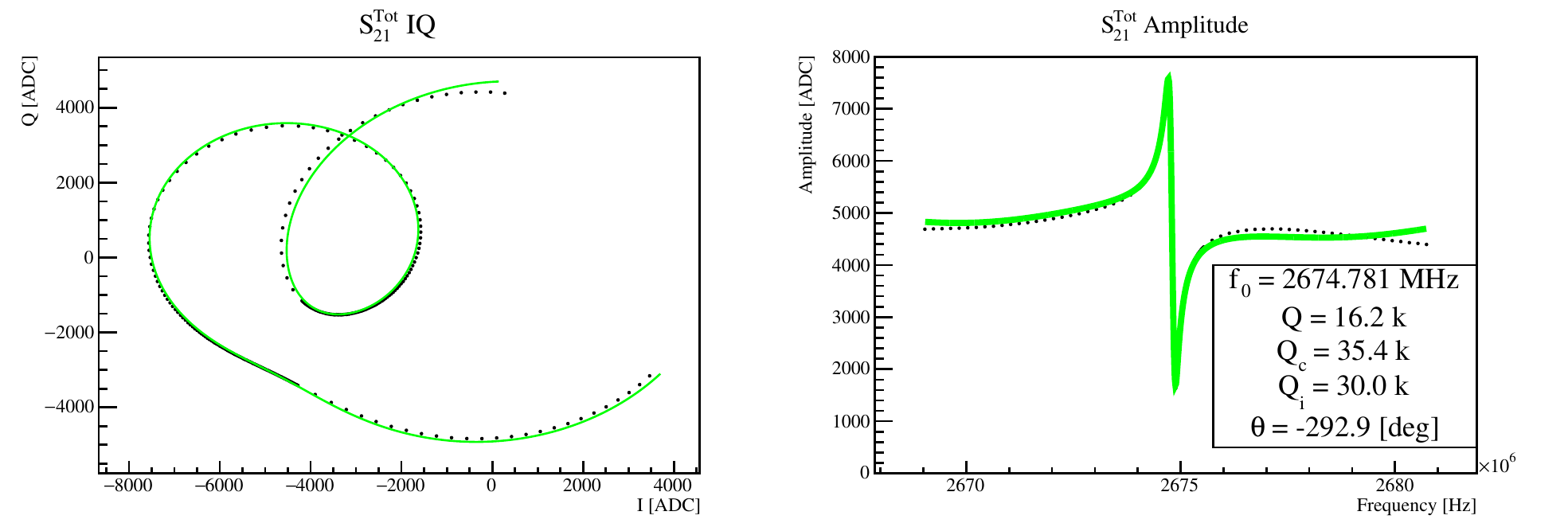}
\end{center}
\caption{{\it Top} panels: Resonance frequency scan data (black bullets) and best fit $S_{21}^{Tot}$ function (green line) in two different representations (IQ plane on the {\it Left} panel, Amplitude vs frequency on {\it Right} panel). The best fit parameters ($f_{0}$, $Q$, $Q_{i}$, $Q_{c}$ and $\theta$) are also given. Another resonance in a different run with a higher impedance mismatch with the feed-line is shown on the {\it Bottom}-{\it Left} (IQ plane) and {\it Bottom}-{\it Right} (Amplitude vs frequency).}
\label{fig1}
\end{figure}
In this work each of these effects is taken into account simultaneously with the following equation:
\begin{equation}
S_{21}^{Tot} = Z_{c} + \left(Acos\left(-2\pi f\tau\right)+j\cdot Bsin\left(-2\pi f\tau\right)\right)\cdot e^{-j\phi} \cdot\left(1-\frac{\frac{\frac{Q}{Q_c}\cdot e^{j\theta}}{1+j2\cdot y}}{\cos{\theta}}\right)
\end{equation}
where 7 parameters describe the feed-line contribution $Z_{c}=(Z_{C_I},Z_{C_Q}$), $A$, $B$, $\tau$, $\phi$ and $\theta$, and 4 parameters the resonator properties $f_{0}$, $Q$, $Q_{c}$ and $a$.\\
The $S_{21}^{Tot}$ complex function was fitted in the (I,Q,$f$) space by minimizing the square modulus of the difference between the function and the data points:
\begin{equation}
min\left(\sum_{n=0}^{N} \parallel S_{21}^{Tot}(f_n,11par)-Data(f_n) \parallel^2\right)
\end{equation}
Fig.~\ref{fig1} shows two examples of the $S_{21}^{Tot}$ function evaluated by fitting one resonance circle of two different chips; in the second one a higher impedance mismatch occurs between the feed-line and the chip: the resonance parameters are estimated despite the huge distortion caused by the mismatch.

\section{Detector working point selection}\label{secWP}
The detectors are operated in the most sensitive point, where the signal-to-noise ratio ($S/N$) is maximum. In order to find this working point each detector is characterized performing a scan of the input power: for each microwave input power the average signal and the average noise power spectrum are acquired. The average pulse is evaluated exploiting the signals produced by the optical system. They are acquired for the real and imaginary parts of each resonator (I and Q) and then converted into changes in phase $\delta\phi$ and amplitude relative to the center of the resonance loop. Since the $\delta\phi$ signal is $\sim6$ times larger than the amplitude one only the $\delta\phi$ signals are used in the analysis.
\begin{figure}
\begin{center}
\includegraphics[%
  width=0.49\linewidth,
  keepaspectratio]{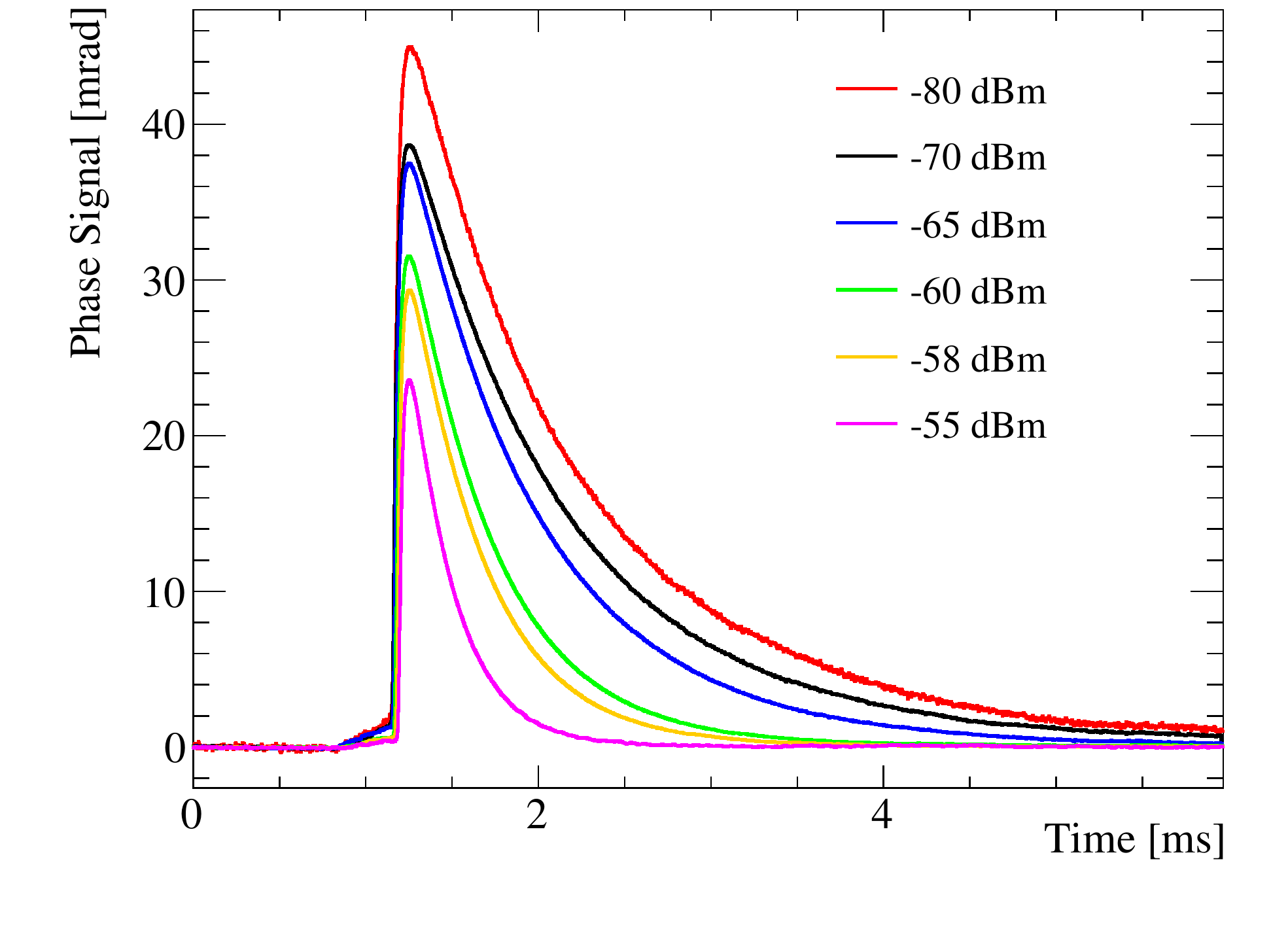}
\includegraphics[%
  width=0.5\linewidth,
  keepaspectratio]{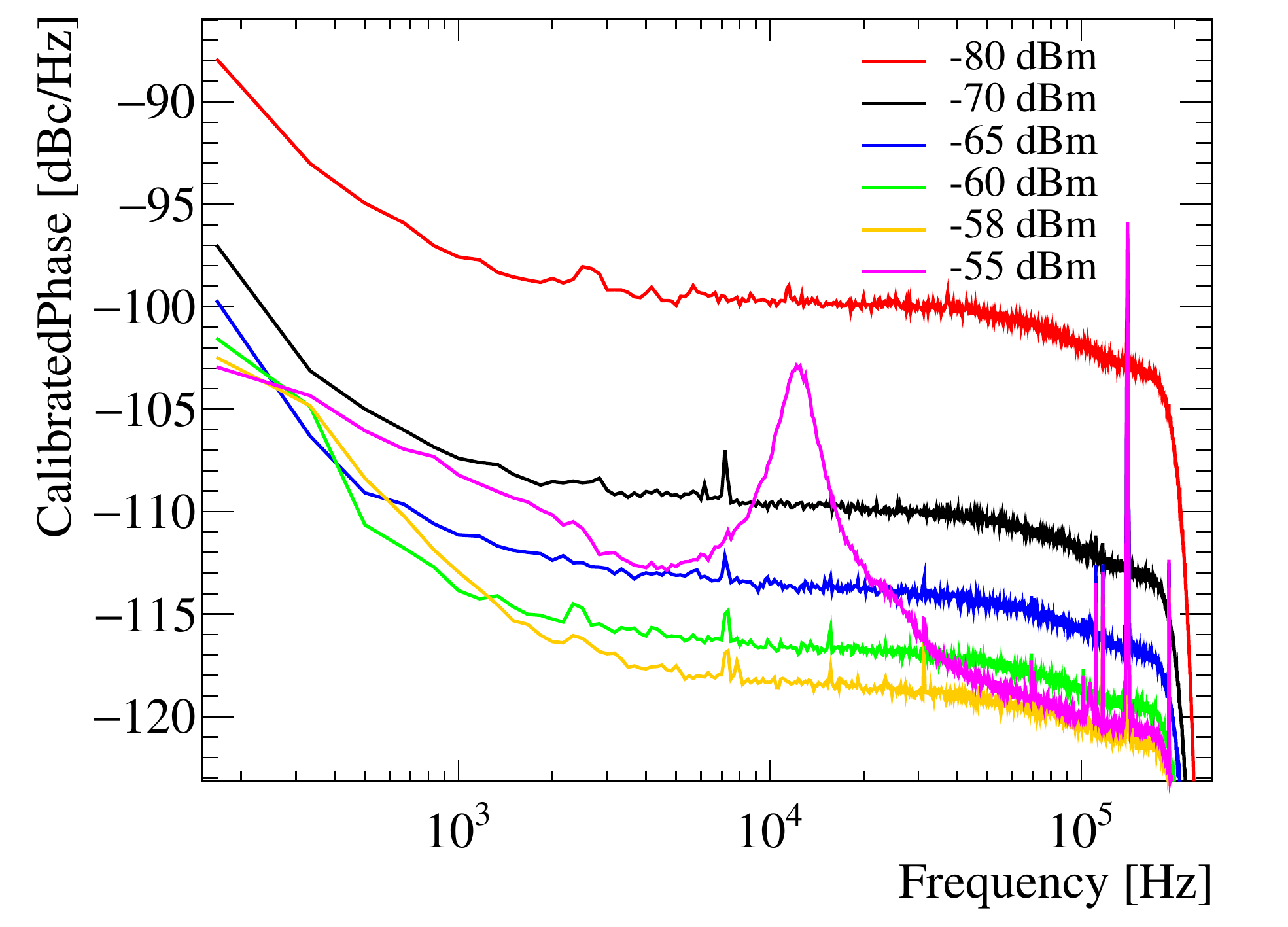}
\end{center}
\caption{{\it Left}: average phase pulse from optical signals for several microwave input power: increasing the power the signal decreases. {\it Right}: average noise power spectrum for several microwave input power: the noise level decrease up to an input power of -58~dBm.}
\label{fig2}
\end{figure}\\
The $\delta\phi$ pulse obtained averaging hundred events produced by optical pulse interactions within the substrate are shown for several microwave input powers in Fig.~\ref{fig2}: since the decay time decreases increasing the microwave input power it has been identified with the quasi-particle life-time $\tau_{qp}$ (see Ref.~\cite{Angelo} for more details). The decrease of $\tau_{qp}$ produces as a consequence a decrease in the signal integration length but, on the other hand, increasing the input power produces a reduction of the noise contribution from the amplifier, as shown in Fig.~\ref{fig2} {\it Right}. The flat noise observed in the high frequency region of the phase read-out is consistent with the noise temperature of the amplifier ($T_{N}\sim􏰁7$~K). The low-frequency region of the phase spectra is dominated by another noise source, whose origin is under investigation. The signal evaluated using the maximum of the average pulses, the noise evaluated integrating the average noise power spectrum filtered with the Optimum Filter algorithm~\cite{OF1,OF2}, and its ratio ($S/N$) as function of the microwave power for one resonator are shown in Fig~\ref{fig3}: for this particular resonator the maximum in $S/N$ is obtained for an input power equal to -58~dBm.
\begin{figure}
\begin{center}
\includegraphics[%
  width=1.\linewidth,
  keepaspectratio]{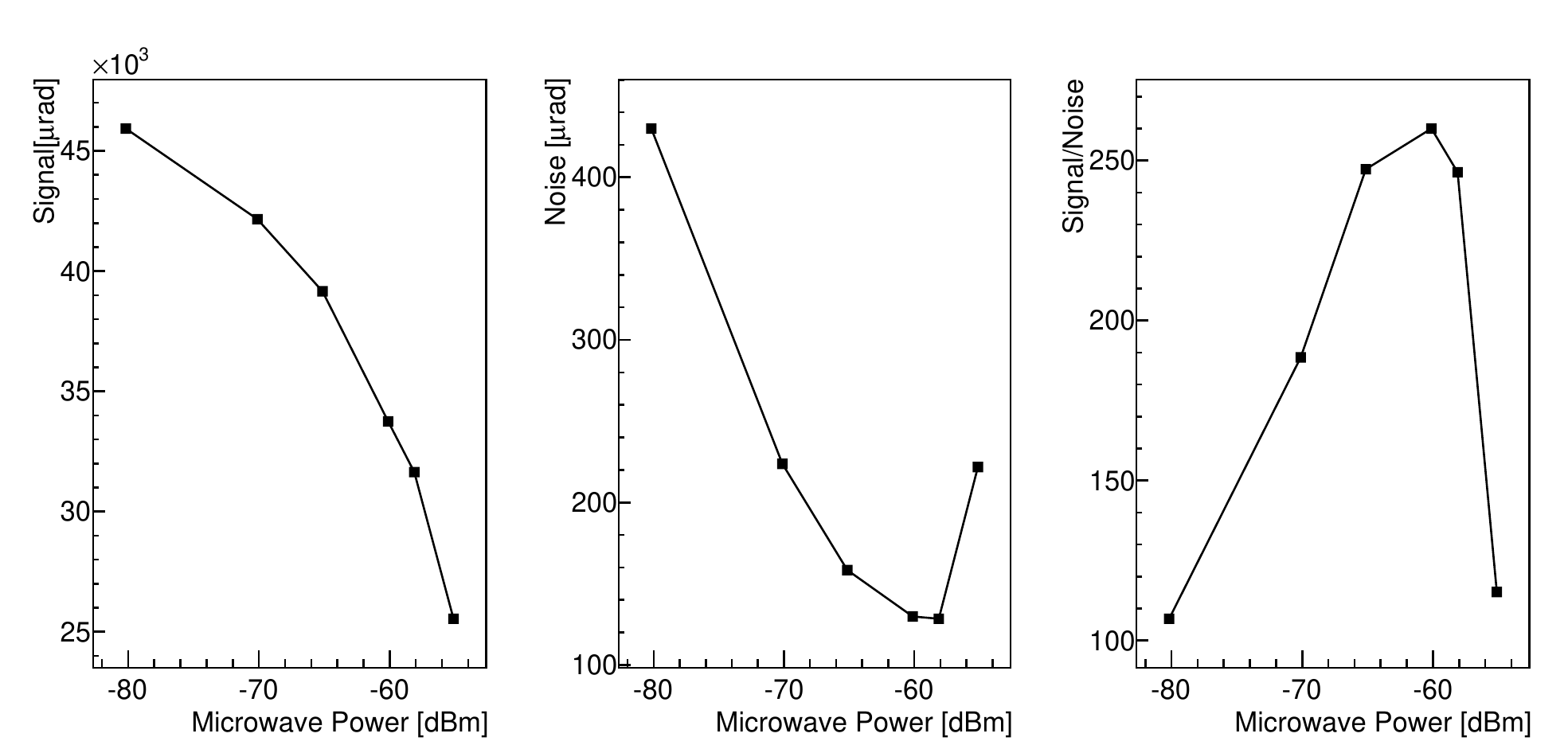}
\end{center}
\caption{Amplitude of the average phase signals {\it Left} and the integrals of the filtered average noise power spectra {\it Center} as function of the microwave input power. The trend of the signal-to-noise ratio ($S/N$) as function of the microwave input power {\it Left}: the maximum of $S/N$ is obtained for an input power of -58~dBm.}
\label{fig3}
\end{figure}

\section{Detector response to optical pulses}
\begin{figure}
\begin{center}
\includegraphics[%
  width=0.6\linewidth,
  keepaspectratio]{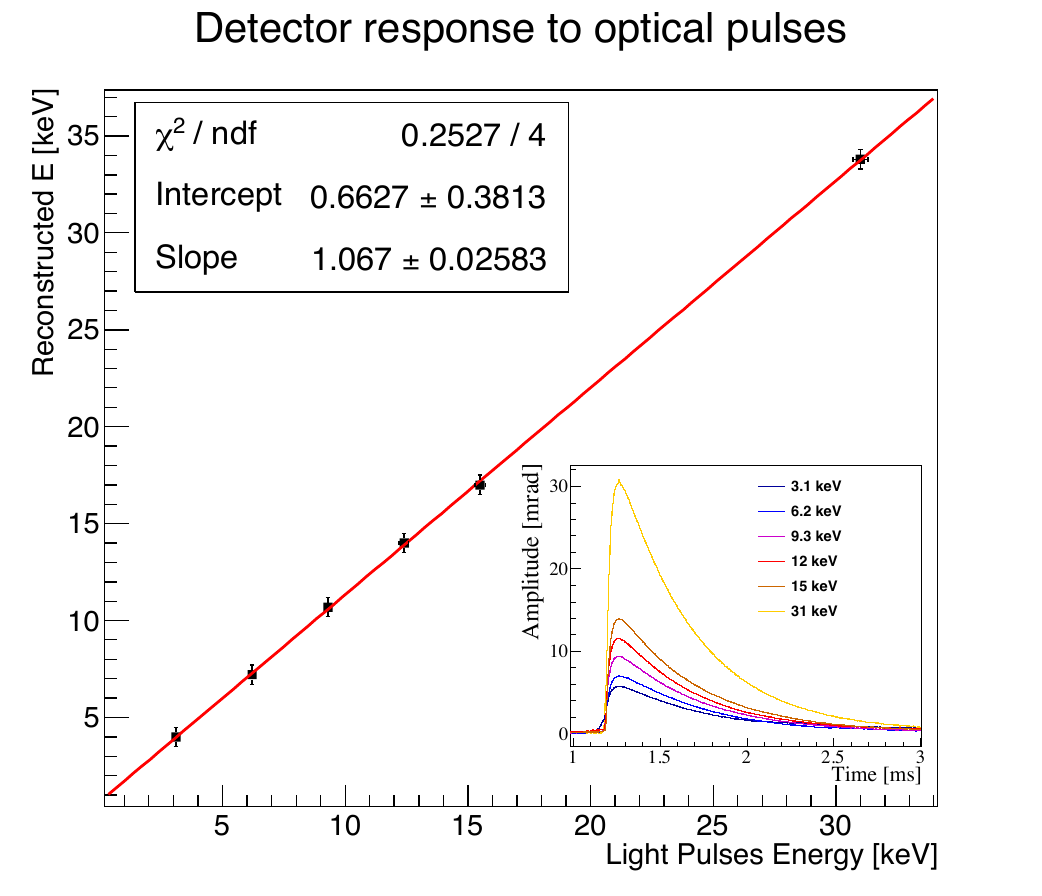}
\end{center}
\caption{Comparison between the expected light pulse energy evaluated applying the room temperature calibration and the Monte Carlo simulation, and the measured one (using the absolute energy calibration process). Data are fitted with a line shape that results with a slope compatible with 1 and an intercept compatible with zero: this result testify the validity of the calibration process also for the optical pulses.}
\label{fig4}
\end{figure}

Since the final purpose is the Cherenkov light detection, the response to 400~nm optical pulses is studied. As discussed in Ref.~\cite{Cardani:2015} the energy absorbed by the resonator and the phase signal integral $\phi_{SigInt}$ are related by a calibration function that allows to perform an absolute energy calibration of the detector. Using that procedure the energy of the incident light pulses can be evaluated and compared with the nominal one: the optical system was previously calibrated at room temperature with a PMT and exploiting a Monte Carlo simulation that takes into account the PMT quantum efficiency, the reflectivity of silicon~\cite{Aspnes}, and the geometrical efficiency of the set-up. The results shown in Fig.~\ref{fig4} demonstrate the validity of the absolute energy calibration process, and at the same time the linear response of the detector for energy depositions up to 30~keV.

\section{Conclusions}
In this work the analysis tools developed for the characterization of the first CALDER aluminum array on Si substrate are presented. A model for the measurement of the resonance parameters able to account for all the effects caused by a non-ideal experimental set-up was developed. The technique for the selection of the optimal detectors point operation (maximum of $S/N$) was discussed. Finally the capability to reconstruct the energy deposition of optical photon pulses with wavelength of 400~nm was demonstrated.

\begin{acknowledgements}
This work was supported by the European Research Council (FP7/2007-2013) under contract CALDER no. 335359 and by the Italian Ministry of Research (FIRB 2012) under contract no. RBFR1269SL.
\end{acknowledgements}



\begin{thebibliography}{99}

\bibitem{Battistelli:2015vha} 
  E.~S.~Battistelli {\it et al.},
  Eur.\ Phys.\ J.\ C {\bf 75}, no. 8, 353 (2015)

\bibitem{Swenson:2010}
  L.~J.~Swenson, {\it et al},
 J.\ Appl.\ Phys.\  {\bf 96} (2010) 263511


\bibitem{Moore:2012}
  D.~C.~Moore, {\it et al},
 J.\ Appl.\ Phys.\  {\bf 100} (2012) 232601


\bibitem{Ivan} 
I.~Colantoni, {\it et al},
  J.\ Low Temp.\ Phys.\ This Special Issue (2015)


\bibitem{Ivan2} 
I.~Colantoni, {\it et al},
Nucl.\ Instrum.\ Meth.\ A, Under Review (2015)

\bibitem{ampli}
http://radiometer.caltech.edu/datasheets/amplifiers/CITLF4.pdf

\bibitem{Bour:2011} 
O.~Bourrion, {\it et al},
  JINST {\bf 6} (2011) P06012

\bibitem{Bour:2013} 
O.~Bourrion, {\it et al},
  JINST {\bf 8} (2013) P12006


\bibitem{Cardani:2015}
L.~Cardani, {\it et al},
Appl.\ Phys.\ Lett. {\bf 107}, (2015) 093508

\bibitem{Khalil:2012}
  M.~S.~Khalil, M.~J.~A.~Stoutimore, F.~C.~Wellstood, and K.~D.~Osborn,
  J.\ Appl.\ Phys.\  {\bf 111} (2012) 054510


\bibitem{Swenson:2013jsa}
  L.~J.~Swenson, {\it et al},
  J.\ Appl.\ Phys.\  {\bf 113} (2013) 104501


\bibitem{Angelo} 
A.~Cruciani, {\it et al},
  J.\ Low Temp.\ Phys.\ This Special Issue (2015)


\bibitem{OF1}
E.~Gatti, and P.~F.~Manfredi,
Nuovo Cimento {\bf 9}, 1, (1986)

\bibitem{OF2}
V.~Radeka, and N.~Karlovac,
Nucl.\ Instrum.\ Methods {\bf 52}, 86 (1967)

\bibitem{Aspnes}
D.~E.~Aspnes and A.~A.~Studna,
Phys.\ Rev.\ B {\bf 27}, (1983) 985.

\end{thebibliography}
\end{document}